# Bright high-repetition-rate source of narrowband extreme-ultraviolet harmonics beyond 22 eV

He Wang[1], Yiming Xu[1], Stefan Ulonska[1], Joseph S. Robinson[1], Predrag Ranitovic[1] & Robert A. Kaindl[1]

*[1]Materials Sciences Division, E. O. Lawrence Berkeley National Laboratory, 1 Cyclotron Road, Berkeley, California 94720, USA*

**Novel table-top sources of extreme-ultraviolet light based on high-harmonic generation yield unique insight into the fundamental properties of molecules, nanomaterials, or correlated solids, and enable advanced applications in imaging or metrology. Extending high-harmonic generation to high repetition rates portends great experimental benefits, yet efficient extreme-ultraviolet conversion of correspondingly weak driving pulses is challenging. Here, we demonstrate a highly-efficient source of femtosecond extreme-ultraviolet pulses at 50-kHz repetition rate, utilizing the ultraviolet second-harmonic focused tightly into Kr gas. In this cascaded scheme, a photon flux beyond $\approx 3 \times 10^{13}$ s$^{-1}$ is generated at 22.3 eV, with $5 \times 10^{-5}$ conversion efficiency that surpasses similar harmonics directly driven by the fundamental by two orders-of-magnitude. The enhancement arises from both wavelength scaling of the atomic dipole and improved spatio-temporal phase-matching, confirmed by simulations. Spectral isolation of a single 72-meV wide harmonic renders this bright, 50-kHz extreme-ultraviolet source a powerful tool for ultrafast photoemission, nanoscale imaging and other applications.**

___________________

Correspondence and requests for materials should be addressed to H.W. (HeWang@lbl.gov) or to R.A.K. (RAKaindl@lbl.gov).

Unique table-top sources of spatially and temporally coherent X-rays are enabled by high-harmonic generation (HHG), which is based on strong-field ionization, acceleration, and recombination of electrons in intense laser fields exceeding $10^{13}$ W/cm².[1–4] Generally, HHG is driven by energetic, mJ-scale lasers operating at low repetition rates – from a few Hz to several kHz – that allow for loose focusing to maximize phase-matching and thus the conversion efficiency.[5,6] Extending tabletop extreme ultraviolet (XUV) sources with ample flux towards rates of ≈50 kHz and beyond is difficult, but can dramatically advance both fundamental investigations of matter and applications in metrology or imaging. For instance, high repetition rates are critical to coincidence and time-of-flight spectroscopy of molecules and solids,[7,8] and can also boost photoemission-based imaging and time-resolved studies where electron space-charge effects require spreading the flux over many pulses.[9–17] Materials studies in particular require a narrow bandwidth ($\lesssim 100$ meV) to discern the electronic structure, and ≈$10^{13}$ photons s⁻¹ HHG source flux before spectral selection for acquisition times comparable to static synchrotron-based photoemission.[14,18] Efficient high repetition-rate HHG, however, is challenging due to the difficulty of phase-matching the conversion in a tight laser focus, necessary to achieve strong-field conditions with μJ-level driving pulses. XUV generation directly from 50–100 kHz Ti:sapphire amplifiers so far resulted in ≈$10^{-8}$ conversion efficiency and a flux up to ≈$3\times10^9$ photons s⁻¹ per harmonic.[19–21]

Different schemes have been pursued to address this challenge, encompassing a large range of approaches that start from high (50-100 kHz) and extend up to ultra-high (multi-MHz) repetition rates. The latter were enabled by increasing the repetition rate thousand-fold via intra-cavity HHG in enhancement resonators, which boosts the average XUV power despite limited (≈$10^{-11}$-$10^{-7}$) conversion efficiencies.[22,23] An optimized setup delivers $10^{12}$-$10^{13}$ photons s⁻¹ harmonic flux around 50 MHz.[24] While ideal for XUV frequency-comb metrology, these oscillator-based schemes are unsuited to ultrafast studies requiring strong excitation pulses and continuous operation is limited by hydrocarbon contamination at kW intra-cavity powers.

Alternatively, high repetition-rate HHG directly with intense driving pulses is enabled by Yb-based solid-state or fiber amplifiers that withstand high average powers.[25–27] At 100 kHz repetition rate, $10^{12}$ photons s$^{-1}$ were generated which corresponds to ≈5·10$^{-7}$ efficiency.[26] Scaling up to 0.6 MHz was achieved by combining the output of multiple fiber amplifiers to 163 W power, yielding more than $10^{13}$ photons s$^{-1}$ flux with up to 2·10$^{-6}$ efficiency.[28] Recently, the absorption limit of infrared (IR) driven HHG was reached using 8-fs pulses and gas pressures up to several bar, yielding broad harmonics with 8·10$^{-6}$ efficiency and ≈2·10$^{12}$ ph s$^{-1}$ flux at 150 kHz.[29] In this context, further enhancement of high repetition-rate HHG motivates the investigation of methods to boost the conversion process itself. Below-threshold harmonics represent one possibility, where phase-matching near atomic resonances enables efficient, yet spectrally broad and structured emission.[30] A second, highly interesting route arises from strong wavelength scaling of the HHG atomic dipole, evidenced at low repetition rates by increased XUV flux resulting from mJ-scale, visible and ultraviolet (UV) pulses loosely focused for optimal phase-matching.[31–34] Efficient HHG with short-wavelength sources, however, has so far not been demonstrated under high repetition-rate conditions, and HHG sources are not typically optimized to generate spectrally narrow harmonics.

Here, we explore UV-driven HHG in the tight-focusing regime and establish a highly-efficient source of narrowband XUV pulses at 50-kHz repetition rate. A bright harmonic flux of ≈3×10$^{13}$ photons s$^{-1}$ is generated at 22.3 eV in this cascaded approach, where efficient frequency doubling is followed by HHG in Kr gas. We establish an XUV conversion efficiency of up to ≈5×10$^{-5}$, which exceeds by two orders-of-magnitude that of similar harmonics driven directly by the near-infrared laser amplifier. This strong boost surpasses the dipole wavelength scaling and, as confirmed by numerical simulations, evidences enhanced spatio-temporal phase-matching for UV-driven harmonics in the sharply focused beam. The spectral structure enables direct isolation of a single, 72-meV wide harmonic – yielding a compact and bright, high repetition-rate XUV source for a new class of ultrafast XUV studies.

# Results

**High-repetition rate XUV generation.** Our scheme is illustrated in Fig. 1a. Near-infrared pulses of 120 µJ energy and 50-fs duration are generated by a cryogenically-cooled, high repetition-rate (50-kHz) Ti:sapphire regenerative amplifier, and focused onto a 0.5-mm thick β–Barium borate (BBO) crystal for frequency doubling with ≈40% efficiency. Here, loose focusing with a $f = 1$ m focal length lens avoids nonlinear spectral broadening and ionization in air. This results in UV pulses centered around 390 nm wavelength with 48 µJ pulse energy, which are separated from the fundamental via two dichroic multilayer mirrors. To initiate the high-harmonic generation process, the femtosecond UV pulses are recollimated and subsequently focused sharply via a $f$=175 mm lens onto a cylindrical gas cell, consisting of an end-sealed glass capillary housed in a vacuum chamber. The capillary is supplied with Kr gas and positioned near the focus of the second-harmonic laser beam, while suppressing the production of off-axis beams arising from long-trajectory electron dynamics.[35]

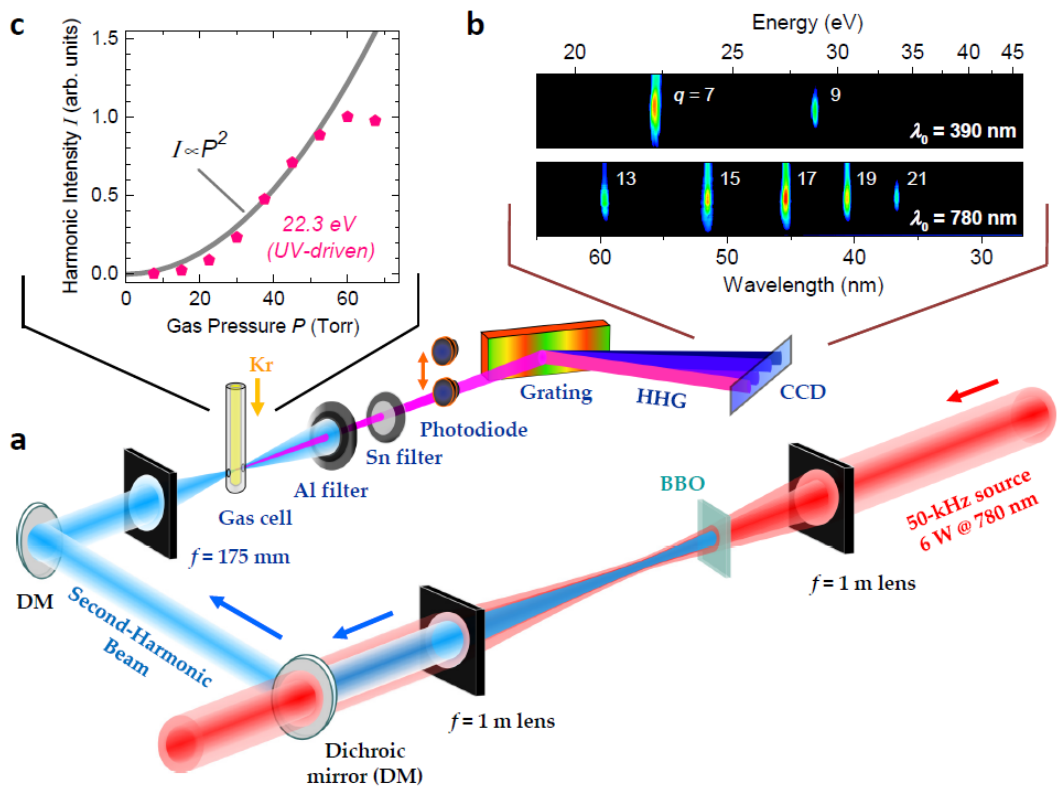

**Figure 1. Efficient high repetition-rate source of extreme ultraviolet (XUV) pulses. a,** Scheme for two-stage high-harmonic generation, starting from 120 µJ near-infrared pulses at 50-kHz repetition rate, which in the first step are frequency doubled to 390 nm wavelength in BBO. These ultraviolet (UV) pulses are subsequently focused sharply onto a thin column of Krypton gas to initiate high-harmonic generation. The resulting XUV light is filtered with thin metal foils, followed by photon flux and spectral characterization with a calibrated XUV photodiode and grating spectrometer. **b,** Intensity-normalized CCD images of the spectrally-dispersed $q$-th XUV harmonics, generated by either the UV pulses or the near-IR fundamental at their respective driving wavelength $\lambda_0$. **c,** Scaling of XUV intensity with Kr gas pressure, for the brightest UV-driven harmonic at 22.3 eV.

Under these conditions we observe the emission of strong XUV harmonics. Their spectral content is characterized by a spectrometer after blocking the diverging optical beam with thin aluminum filters (see Methods). Figure 1b (top image) shows the resulting charge-coupled device (CCD) readout for a gas pressure of 60 Torr. Bright peaks are detected around 22.3 eV and 28.6 eV, respectively corresponding to the 7$^{th}$ and 9$^{th}$ harmonics of the UV driving pulses. For direct comparison, we also recorded the XUV emission generated by directly focusing the ≈780 nm fundamental pulses onto the Kr gas to a similar peak intensity. Multiple harmonics are observed, as shown in the corresponding normalized CCD data in Fig. 1b (bottom image). However, the emission was found to be significantly weaker compared to the UV-driven harmonics.

Figure 1c shows the dependence of the XUV intensity of the most intense harmonic on the backing pressure of the Kr gas target, indicating a rapid nonlinear increase up to about 60 Torr followed by saturation at higher pressures. The strong XUV emission arises from phase-matched harmonic generation, due to in-phase coherent addition of the XUV fields emitted from the Kr atoms. At higher pressures plasma defocusing limits the overall yield.[20,36,37] We have also generated harmonics in Ar gas for comparison, resulting however in ≈5 times lower yield. For optimal phase-matching in Kr, the vertical intensity contour on the CCD image (Fig. 1b) perpendicular to the dispersion direction indicates an approximately Gaussian beam profile.

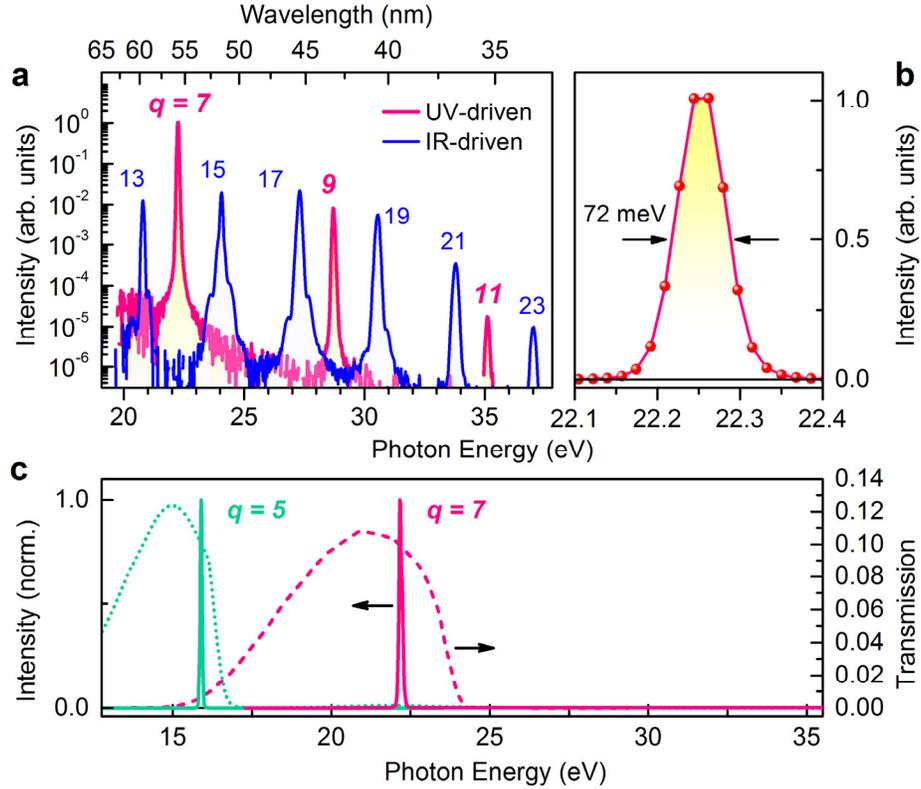

**Figure 2. XUV spectra and isolation of a single harmonic. a,** Spectra of the XUV harmonics (vertical CCD lineouts) driven either by the 48-μJ UV pulses or by the 120-μJ pulses from the near-IR laser fundamental. Spectra are measured for a Kr gas pressure of 60 Torr, and are shown for the same integration time of 4 s and with CCD background noise subtracted. For the UV-driven harmonics, the spectrum was corrected for the effect of one additional Al filter, inserted to avoid CCD saturation. **b,** Emission profile of the 7$^{th}$ harmonic at 22.25 eV with corresponding line width (FWHM). **c,** Isolated single harmonics (solid lines) after absorptive spectral filtering with thin metal foils. The 7$^{th}$-harmonic is isolated via combined Sn and Al filters of 300-nm thickness each, while the 5$^{th}$-harmonic is selected by an In foil in combination with Brewster reflection from two Si plates[38] to suppress the residual laser beam. For comparison, the theoretical transmission is shown for Sn (dashed line) and In (dotted line) of 300-nm thickness.[49]

From its extent, we obtain a beam divergence of 6 mrad (full width at half maximum, FWHM), which facilitates extended beam propagation for refocusing or additional optical manipulations.

**Spectral structure and harmonic flux.** The experiments reveal a striking enhancement of the XUV emission generated by the 390-nm field as compared to that obtained from the near-IR fundamental. Figure 2a compares the measured XUV spectra for the same integration time, corrected for an additional Al filter used to avoid CCD saturation. The spectra encompass a series of odd harmonics that span photon energies up to ≈37 eV with strongly varying peak intensities, where the 7$^{th}$ harmonic of the UV field constitutes the by far strongest XUV emission. The latter, UV-driven peak at ≈22.3 eV surpasses that of the spectrally-similar 13$^{th}$ and 15$^{th}$ harmonics of the near-IR fundamental by 50–80 times. This strong enhancement is particularly striking, as it occurs despite the necessarily lower energy of the UV pulses derived from frequency doubling. Importantly, this harmonic exhibits a linewidth as narrow as 72 meV FWHM near optimum pressure, as shown in Fig. 2b. This yields a critical advantage for applications such as photoelectron spectroscopy or zone-plate imaging.

The UV-driven XUV generation entails a large energy separation of ≈6.4 eV between the individual odd harmonics. We can exploit this distinctive feature to isolate a single harmonic from the comb, taking advantage of the atomic and plasma absorption edges of thin metal foils. To select harmonics around 22 eV the beam is passed through 300-nm thick Sn, whose transmission is indicated by the dashed line in Fig. 2c and which strongly attenuates the adjacent 5$^{th}$ and 9$^{th}$ harmonics. Together with the Al foil used to block the residual visible beam, for a combined XUV transmission of ≈1%, we obtain the spectrum in Fig. 2c (magenta line). In this scheme, the 5$^{th}$ and 3$^{rd}$ harmonics at lower energies are also suppressed via the Al foil that blocks the ultraviolet driving field. By attenuating the laser beam instead via two Brewster's-angle reflections from silicon,[38] the lower-order harmonics are transmitted. The 5$^{th}$ harmonic at 15.9 eV is then spectrally selected using a 300-nm thick In foil, as demonstrated in Fig. 2c (green line), with a flux and bandwidth comparable to the isolated $q = 7$ harmonic. Thus, high-contrast isolation of individual UV-driven harmonics is achieved at either 15.9 or 22.3 eV, resulting in a compact source of narrowband and single-harmonic XUV pulses without the need for a complex monochromator.

To obtain the absolute XUV photon flux for each harmonic, the total source power was determined with a calibrated X-ray photodiode, taking into account the filter transmission, and then split according to the harmonic ratios (see Methods). Table 1 shows the photon flux of the individual harmonics and resulting energy conversion efficiencies. Up to $3.3\times10^{13}$ photons s$^{-1}$ are generated in the 7$^{th}$ harmonic of the UV driving pulses, corresponding to 117 µW source power emitted from the Kr gas at 22.3 eV. The corresponding HHG efficiency ($5\times10^{-5}$) is ≈140–400 times higher than for the spectrally closeby 15$^{th}$ and 13$^{th}$ harmonics of the fundamental. The values must be compared at similar photon energies due to the wavelength-dependent opacity of the gas.[39] By contrast, when the influence of phase-matching can be neglected, a scaling $\propto \lambda_0^{-4.7\pm1}$ with visible driver wavelength was previously found,[40] which contributes a ≈12 to 52-fold increase upon frequency doubling of $\lambda_0$. This comparison underscores the significant phase-matching advantage of the UV-driven HHG under our experimental conditions.

**Phase-matching simulations of UV-driven HHG.** In the following, we present numerical HHG simulations to clarify the phase-matching boost of UV-driven HHG with our experimental parameters. The generated number of XUV photons in the $q$-th harmonic can be expressed as[37] $N_q \propto \beta_S(q,\omega_0) \times \xi_q$. Here, $\beta_S$ is the single-atom efficiency due to electronic wavepacket dynamics, which approximately follows $\beta_S \propto \omega_0^5$ around the cutoff due to wavepacket quantum diffusion and energy scaling.[41] In turn, $\xi_q$ is the enhancement factor arising from the spatio-temporal folding of XUV emission and phase-matching.[5,37] In the simulation, the XUV flux generated from each temporal slice of the driving field results from the coherent superposition of all harmonic emissions, integrated across the interaction volume.[5] For this, we take into account the XUV generation at each location $z$, reabsorption by the Kr gas, ground state depletion, and the wavevector mismatch between the high-harmonic and driving fields

$$\Delta k(z,t) \equiv [1-\eta(z,t)] \cdot \Delta k_N + \eta(z,t) \cdot \Delta k_P + \Delta k_D(z,t) + \Delta k_G(z) \tag{1}$$

**Table 1. Source photon flux and energy conversion efficiency per harmonic.**

| Harmonic order | Photon energy (eV) | Photon flux (ph s$^{-1}$) | Conversion efficiency |
|---|---|---|---|
| *Near-IR driving pulse ($\lambda$ = 780 nm)* | | | |
| 13$^{th}$ | 20.8 | $2.2\times10^{11}$ | $1.2\times10^{-7}$ |
| 15$^{th}$ | 24 | $5.3\times10^{11}$ | $3\times10^{-7}$ |
| 17$^{th}$ | 27.3 | $1.6\times10^{12}$ | $1.2\times10^{-6}$ |
| 19$^{th}$ | 30.5 | $1.5\times10^{12}$ | $1.2\times10^{-6}$ |
| 21$^{st}$ | 33.7 | $1.5\times10^{11}$ | $1.3\times10^{-7}$ |
| 23$^{rd}$ | 37 | $1.9\times10^{9}$ | $2\times10^{-9}$ |
| *Ultraviolet driving pulse ($\lambda$ = 390 nm)* | | | |
| 7$^{th}$ | 22.3 | $3.3\times10^{13}$ | $5\times10^{-5}$ |
| 9$^{th}$ | 28.7 | $1.7\times10^{12}$ | $3\times10^{-6}$ |
| 11$^{th}$ | 35 | $8.3\times10^{9}$ | $2\times10^{-8}$ |

The first two terms above describe the mismatch due to dispersion at a given pressure, where $\Delta k_N$ corresponds to the charge-neutral gas and $\Delta k_P$ to a fully ionized plasma, accordingly scaled by the ionization level $\eta$. Moreover, $\Delta k_D$ results from the atomic dipole phase accumulated by the accelerated electron wavepacket, and $\Delta k_G$ from the geometric Gouy phase of the focused beams.

For loose focusing attainable with mJ pulses, the spatial dependence and Gouy-phase contribution are negligible. Phase-matching is then achieved at suitable pressures by dynamically balancing to $\Delta k \approx 0$ via the ionization level, given that the sign of the plasma contribution ($\Delta k_P > 0$) is opposite to $\Delta k_N$ of neutral atoms. Such conditions are fulfilled only during part of the driving pulse, since the ionization level rises quickly with time.[3] Instead, under tight focusing where the Rayleigh length $z_R$ is comparable to the gas cell thickness, the conversion efficiency is reduced due to the added spatial dependence of $\Delta k$. The Gouy term now becomes significant and as its sign equals the plasma contribution, phase-matching occurs at a lower ionized fraction and thus reduced intensity. Although increased pressures can partly compensate this reduction, the tolerable gas density is limited by plasma defocusing and pumping capabilities.[36,37,42]

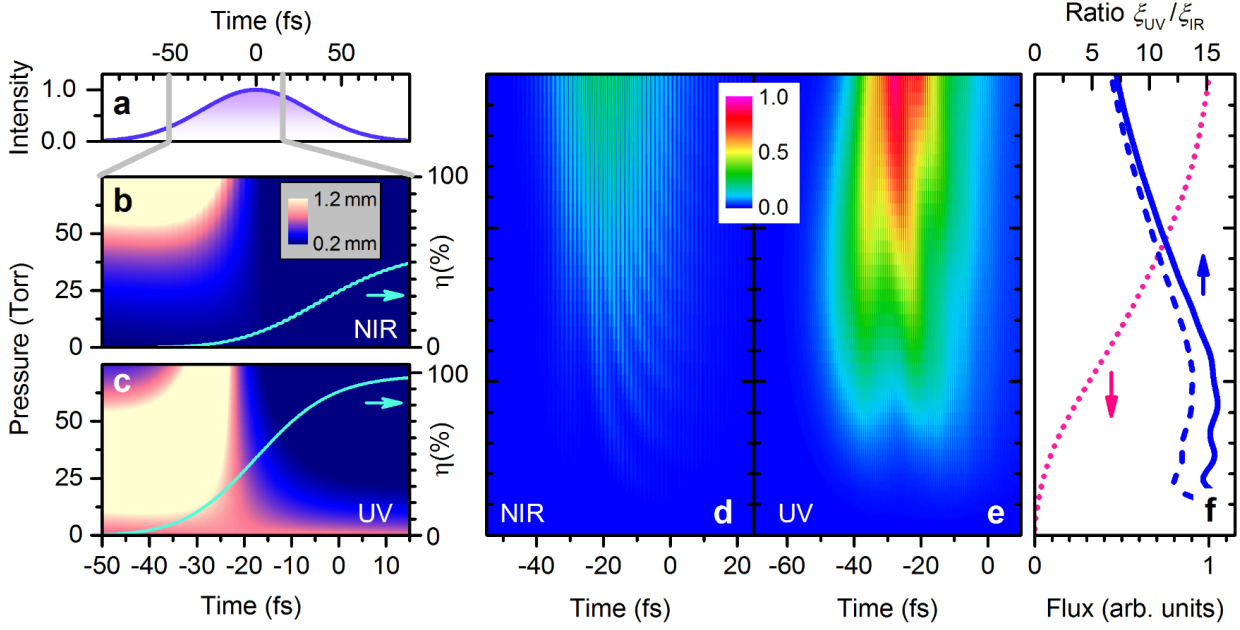

**Figure 3. HHG phase matching simulations in the tight focus geometry.** The calculations are for a 75-fs FWHM driving pulse with $1.8 \cdot 10^{14}$ W/cm$^2$ peak intensity, Rayleigh length of $z_R = 1$ mm, and with the gas volume centered at $z = z_R/2$ after the focus. **a,** Driving pulse intensity. **b,c,** Coherence length $\pi/|\Delta k|$, mapped as function of time and Kr gas pressure, comparing the phase-matching at similar XUV photon energies of the near-IR (NIR) driven 15$^{th}$ and UV-driven 7$^{th}$ harmonic. Plots are shown at 200 μm before the gas volume exit, comparable to the XUV absorption length at 50 Torr. The Kr ionization level is shown for comparison (blue line) as calculated with the YI model.[43] **d,e,** Resulting XUV photon flux enhancement, obtained by spatially integrating the HHG emission across the interaction volume (see Supplementary Methods) for the two cases, with the single-atom efficiency $\beta_s$ omitted. The flux enhancement is mapped as a function of gas pressure and emission time within the driving pulse. **f,** Pressure dependence of the UV-driven XUV emission (dotted line), obtained by integrating the data in panel e across the pulse. The ratio of the time-integrated phase-matching enhancements $\xi_{UV}/\xi_{IR}$ is shown for the above model (dashed line), and for better comparison with experiment using a 50-fs near-IR driving pulse (solid line).

When UV driving pulses are employed, the above spatio-temporal effects of phase-matching on the XUV emission – represented by the enhancement $\xi_q$ – are substantially improved. We have calculated phase-matched HHG in Kr gas with the above model under our conditions, for pulses with either 390 nm or 780 nm center wavelength and otherwise identical $1.8 \times 10^{14}$ W/cm$^2$ peak intensity and 75-fs duration (for details see Supplementary Methods). To mimic the experiment, we further consider a Rayleigh length of $z_R = 1$ mm and a 1-mm thick gas volume. The driving pulse profile is indicated in Fig. 3a. Figures 3b and 3c map the coherence length $L_{coh} = \pi/|\Delta k|$ as a function of gas pressure and time. The Kr ionization levels are shown for comparison (lines in Figs 3b and 3c) and ramp up quickly towards the pulse center. They are calculated with the Yudin-Ivanov model[43] to encompass both tunneling and multi-photon ionization, whose relative influence was illustrated in previous studies of HHG phase-matching and wavelength scaling under loose focusing.[44] Several aspects contribute to the enhanced phase-matching in the UV field. First, due to $1/\omega_0^2$ scaling of the plasma mismatch, higher ionization levels are tolerated in the UV as evident from a comparison of Figs. 3b and 3c. The ionization rates are accordingly higher, and our calculations indicate a $\approx$4.2 times enhancement over IR-driven HHG. Secondly, the spatial phase matching conditions are also improved, since the Gouy phase is proportional to the harmonic order. As a result, the absolute value and spatial dependence of $\Delta k_G$ are reduced by half when driven by the second-harmonic. For these UV-driven enhancements we note that plasma scaling equally applies to loose focusing, while the Gouy phase reduction only matters for tight focusing entailed by limited pulse energies of high repetition-rate lasers.

To obtain the total XUV flux we integrate the harmonic emission across the gas volume, taking into account the gas density, XUV emission and re-absorption, and the spatio-temporal dynamics of the ionization level and phase mismatch. Figures 3d and 3e compare the resulting XUV flux enhancement beyond the single-atom efficiency $\beta$s, for the two incident wavelengths and mapped out for different pressures and temporal slices within the driving field. This illustrates the time window of XUV emission and demonstrates the clear phase-matching advantage of UV-driven HHG. Figure 3f shows the pressure dependence of the integrated XUV flux generated by the UV field (dotted line). The ratio of the total, time-integrated phase-matching enhancements $\xi_q(UV)/\xi_q(IR)$ is shown in Fig. 3f (dashed line). Further improvement is found when considering the shorter $\approx$50-fs IR pulses in the experiment (solid line), resulting in a 10-16 times UV-driven boost of phase-matched XUV generation below 60 Torr. Combined with the $\approx\omega_0^5$ scaling of the single-atom efficiency that contributes an additional factor of 32, this UV-driven boost of phase-matched XUV generation explains the two orders-of-magnitude increase in HHG efficiency observed in our experiments.

**Discussion**

The efficient HHG conversion demonstrated here at 50-kHz results in a unique XUV source with major experimental benefits. Beyond enhancing signal averaging and statistics in numerous applications, high repetition rates are particularly valuable to advancing photoemission electron microscopy[17] or angle-resolved photoelectron spectroscopy (ARPES),[11,13] where space-charge Coulomb interactions between the emitted electrons limit acceptable pulse energies. Moreover, in ARPES the accessible momentum space increases rapidly with photon energy, and XUV photons beyond 20 eV render the full Brillouin zone of most materials easily accessible.

After high-contrast spectral isolation of the 22.3 eV harmonic with the combined Al-Sn filter, our femtosecond source delivers $\approx 3\times 10^{11}$ photons s$^{-1}$ at the sample. This flux is comparable to continously-emitting Helium lamps and even approaches that of monochromatized synchrotron beamlines employed for static ARPES.[14,18] As space-charge broadening typically restricts each pulse to $\approx 10^6$ photons, the 50-kHz repetition rate at this flux provides ideal conditions for photoemission studies.[14,45] We note that by utilizing only the Al foil, even higher flux beyond $\approx 3\times 10^{12}$ photons s$^{-1}$ is obtained at the expense of isolation contrast. The addition of the Sn filter, however, also provides a compact way to separate the gas-based XUV source and attached optics chambers from a subsequent ultrahigh-vacuum environment.

Spectral isolation and narrowing of high-order harmonics is often achieved using XUV monochromators, which add significant complexity and require non-traditional layouts for time-resolved applications to minimize grating-induced pulse broadening.[46,47] In contrast, the UV-driven HHG source reported here demonstrates the direct generation of narrow harmonics, whose large energy spacing allows for straightforward selection with absorptive metal filters. The measured 72-meV harmonic width corresponds to a $\approx 0.3\%$ fractional energy bandwidth directly from the source. This intrinsic resolution is well adapted to discerning the electronic structure of atoms, molecules, and complex materials via photoemission studies.

Hence, XUV conversion with high efficiency is achieved at 50-kHz repetition rate, resulting from both wavelength scaling of the atomic dipole and enhanced phase-matching conditions in a UV field under tight focusing conditions. While based on Ti:sapphire amplified pulses, this cascaded scheme is generally applicable and attractive for enhancing HHG also with other high-repetition rate sources. In particular, combining UV-driven HHG with novel high-power solid-state[48] or Yb-fiber amplifiers[25,26,28] may help scale XUV generation to even higher flux and repetition rates. Finally, additional narrowing of the harmonics would propel scientific studies of low-energy correlations in solids, which motivates further HHG studies with longer or spectrally-shaped UV fields. We expect that the compact 50-kHz source of bright and narrowband XUV harmonics, established here, will boost applications in photoemission and nanoscale imaging, paving the way for novel insights into complex matter.

## Methods

**Femtosecond UV Driving Pulses.** The initial stage of the setup is a high average power cryo-cooled Ti:sapphire regenerative amplifier (KMLabs Wyvern 500) that provides near-IR pulses of 50-fs duration at 50-kHz repetition rate, with a high beam quality ($M^2$=1.3× diffraction limited). After splitting off half of the output for photo-excitation in time-resolved applications, the remaining pulses with ≈120 μJ energy were focused onto a 0.5-mm thick BBO crystal with a $f$ = 1 m lens. The crystal is cut for phase-matched second-harmonic generation ($\theta$ = 29.2°) and positioned 380 mm before the focus, corresponding to a peak intensity of ≈100 GW/cm$^2$. We estimate a UV pulse duration of ≈73 fs by taking into account nonlinear propagation in BBO (using the code SNLO) and pulse dispersion in the focusing optics. Pulses with 3 nm bandwidth are generated around 390 nm wavelength with ≈39.5% conversion efficiency.

**Generation and Spectral Characterization of XUV Harmonics.** For high-harmonic generation, Kr gas is introduced into the end-sealed glass capillary positioned within a vacuum chamber. Here, a 1-mm inner diameter and 100-μm wall thickness of the capillary was chosen to accomodate the $z_R$ ≈ 1 mm Rayleigh length of the driving beam. Before use, beam entry and exit holes are laser-drilled in situ at ambient pressure, perpendicular to the capillary side walls via the femtosecond UV laser pulses. Under normal operation, a 750 L·s$^{-1}$ turbopump maintains the pressure in the HHG chamber below 1 mTorr. The capillary backing pressure is monitored with a Si diaphragm gauge.

For XUV generation, the UV pulses are re-collimated with a $f$ = 1-m fused silica lens and then focused onto the gas cell with a $f$ = 175 mm lens, resulting in a 18 μm FWHM beam diameter and $I_0$ ≈ 1.8×10$^{14}$ W/cm$^2$ peak intensity. Using the fundamental for harmonic generation, these values are 33 μm and $I_0$ ≈ 1.9×10$^{14}$ W/cm$^2$. After the generation chamber, the intense driving beam is blocked and the XUV harmonics are spectrally selected via thin metal filters, mounted on two gate valves to enable their insertion. The XUV spectra are recorded with an evacuated spectrometer (McPherson 234/302) equipped with a back-illuminated X-ray CCD and a 2400 l·mm$^{-1}$ Pt-coated aberration-corrected concave grating. The CCD covers the 25-75 nm range ($\approx$49.6–16.5 eV) at 50 nm central wavelength, and a 5-µm wide entrance slit is used for $\approx$0.1 nm spectral resolution.

**XUV Photon Flux.** The absolute photon flux for each harmonic is quantitatively determined by first recording the spectrally-integrated XUV flux with a Si photodiode calibrated in this range (IRD AXUV100G), and then splitting this total flux according to the relative ratios of the harmonics. Residual leakage at the drive wavelength is removed by subtracting the background measured without the noble gas. The total charge collected by the photodiode is the sum of contributions from each $q$-th odd harmonic, i.e.

$$Q_{\text{XUV}} = \sum_q E_q \, T_f(\omega_q) R_{\text{AXUV}}(\omega_q), \qquad (2)$$

where $E_q$ is the pulse energy within the $q$-th harmonic as emitted at the source, $\omega_q$ is the harmonic frequency, $T_f(\omega)$ is the Al filter transmission, and $R_{\text{AXUV}}(\omega)$ the calibrated photodiode responsivity. The Al filter was calibrated in situ by measuring the transmission of near-IR driven harmonic peaks on the X-ray CCD (Supplementary Fig. 1), with the foil inserted with a gate valve. The absolute values of $E_q$ are then obtained from $Q_{\text{XUV}}$ using the relative harmonic ratios. These ratios are determined from the CCD spectrum of the XUV harmonics, corrected for filter transmission and grating efficiency, i.e. from the count rate

$$N_q^{\text{CCD}} \propto E_q \, T_f(\omega_q)^n \eta_{\text{gr}}(\omega_q) QE_{\text{CCD}}(\omega_q), \qquad (3)$$

where $\eta_{\text{gr}}(\omega)$ is the grating diffraction efficiency provided by the manufacturer and $QE_{\text{CCD}}(\omega)$ is the CCD quantum efficiency. To avoid saturation of the CCD camera, the XUV flux was suppressed with either $n = 1$ or 2 filters for the IR and UV-driven harmonics, respectively. The above procedure thus provides the pulse energy and the photon flux $E_q/\hbar\omega_q$ for each harmonic.

# References

1. Corkum, P. B. Plasma perspective on strong-field multiphoton ionization. *Phys Rev. Lett.* **71,** 1994–1997 (1993).

2. Popmintchev, T., Chen, M.-C., Arpin, P., Murnane, M. M. & Kapteyn, H. C. The attosecond nonlinear optics of bright coherent x-ray generation. *Nat. Photonics* **4,** 822–832 (2010).

3. Chang, Z. *Fundamentals of Attosecond Optics*. CRC Press, Boca Raton (2011).

4. McPherson, A. *et al.* Studies of multiphoton production of vacuum-ultraviolet radiation in the rare gases. *J. Opt. Soc. Am B* **4,** 595–601 (1987).

5. Constant, E. *et al.* Optimizing high harmonic generation in absorbing gases: model and experiment. *Phys. Rev. Lett.* **82,** 1668–1671 (1999).

6. Takahashi, E., Nabekawa, Y. & Midorikawa, K. Generation of 10-μJ coherent extreme-ultraviolet light by use of high-order harmonics. *Opt. Lett.* **27,** 1920–1922 (2002).

7. Sandhu, A. S. *et al.* Observing the creation of electronic Feshbach resonances in soft x-ray–induced $O_2$ dissociation. *Science* **322,** 1081–1085 (2008).

8. Gotlieb, K., Hussain, Z., Bostwick, A., Lanzara, A. & Jozwiak, C. Rapid high-resolution spin- and angle-resolved photoemission spectroscopy with pulsed laser source and time-of-flight spectrometer. *Rev. Sci. Instrum.* **84,** 093904 (2013).

9. Haight, R. Electron dynamics at surfaces. *Surf. Sci. Rep.* **21,** 275–325 (1995).

10. Schmitt, F. *et al.* Transient electronic structure and melting of a charge density wave in $TbTe_3$. *Science* **321,** 1649–1652 (2008).

11. Rohwer, T. *et al.* Collapse of long-range charge order tracked by time-resolved photoemission at high momenta. *Nature* **471,** 490–493 (2011).

12. Graf, J. *et al.* Nodal quasiparticle meltdown in ultrahigh-resolution pump–probe angle-resolved photoemission. *Nat. Phys.* **7,** 805–809 (2011).

13. Gierz, I. *et al.* Snapshots of non-equilibrium Dirac carrier distributions in graphene. *Nat. Mater.* **12,** 1119–1124 (2013).

14. Zhou, X. J. *et al.* Space charge effect and mirror charge effect in photoemission spectroscopy. *J. Electron Spectros. Relat. Phenomena* **142,** 27–38 (2005).

15. Siefermann, K. R. *et al.* Binding energies, lifetimes and implications of bulk and interface solvated electrons in water. *Nat. Chem.* **2,** 274–279 (2010).

16. Suzuki, T. Time-resolved photoelectron spectroscopy of non-adiabatic electronic dynamics in gas and liquid phases. *Int. Rev. Phys. Chem.* **31,** 265–318 (2012).

17. Chew, S. H. *et al.* Time-of-flight-photoelectron emission microscopy on plasmonic structures using attosecond extreme ultraviolet pulses. *Appl. Phys. Lett.* **100,** 051904 (2012).

18. Damascelli, A., Hussain, Z. & Shen, Z.-X. Angle-resolved photoemission studies of the cuprate superconductors. *Rev. Mod. Phys.* **75,** 473–541 (2003).

19. Lindner, F. *et al.* High-order harmonic generation at a repetition rate of 100 kHz. *Phys. Rev. A* **68,** 013814 (2003).
20. Chen, M.-C. *et al.* Spatially coherent, phase matched, high-order harmonic EUV beams at 50 kHz. *Opt. Express* **17,** 17376–17383 (2009).
21. Heyl, C. M., Güdde, J., L'Huillier, A. & Höfer, U. High-order harmonic generation with µJ laser pulses at high repetition rates. *J. Phys. B At. Mol. Opt. Phys.* **45,** 074020 (2012).
22. Cingöz, A. *et al.* Direct frequency comb spectroscopy in the extreme ultraviolet. *Nature* **482,** 68–71 (2012).
23. Bernhardt, B. *et al.* Vacuum ultraviolet frequency combs generated by a femtosecond enhancement cavity in the visible. *Opt. Lett.* **37,** 503–505 (2012).
24. Lee, J., Carlson, D. R. & Jones, R. J. Optimizing intracavity high harmonic generation for XUV fs frequency combs. *Opt. Express* **19,** 23315–23326 (2011).
25. Hädrich, S. *et al.* Generation of µW level plateau harmonics at high repetition rate. *Opt. Express* **19,** 19374–19383 (2011).
26. Cabasse, A., Machinet, G., Dubrouil, A., Cormier, E. & Constant, E. Optimization and phase matching of harmonic generation at high repetition rate. *Opt. Lett.* **37,** 4618–4620 (2012).
27. Vernaleken, A. *et al.* Single-pass high-harmonic generation at 20.8 MHz repetition rate. *Opt. Lett.* **36,** 3428–3430 (2011).
28. Hädrich, S. *et al.* High photon flux table-top coherent extreme-ultraviolet source. *Nat. Photonics* **8,** 779–783 (2014).
29. Rothhardt, J. *et al.* Absorption-limited and phase-matched high harmonic generation in the tight focusing regime. *New J. Phys.* **16,** 033022 (2014).
30. Chini, M. *et al.* Coherent phase-matched VUV generation by field-controlled bound states. *Nat. Photonics* **8,** 437–441 (2014).
31. Ditmire, T., Crane, J. K., Nguyen, H., DaSilva, L. B. & Perry, M. D. Energy-yield and conversion efficiency measurements of high-order harmonic radiation. *Phys. Rev. A* **51,** 902–905 (1995).
32. Kim, I. *et al.* Highly efficient high-harmonic generation in an orthogonally polarized two-color laser field. *Phys. Rev. Lett.* **94,** 243901 (2005).
33. Shiner, A. D. *et al.* Wavelength scaling of high harmonic generation efficiency. *Phys. Rev. Lett.* **103,** 073902 (2009).
34. Falcão-Filho, E. L. *et al.* Scaling of high-order harmonic efficiencies with visible wavelength drivers: A route to efficient extreme ultraviolet sources. *Appl. Phys. Lett.* **97,** 061107 (2010).
35. Salieres, P., L'Huillier, A. & Lewenstein, M. Coherence control of high-order harmonics. *Phys. Rev. Lett.* **74,** 3776–3779 (1995).
36. Altucci, C. *et al.* Influence of atomic density in high-order harmonic generation. *J. Opt. Soc. Am. B* **13,** 148–156 (1996).

37. Lai, C.-J. & Kärtner, F. X. The influence of plasma defocusing in high harmonic generation. *Opt. Express* **19,** 22377–22387 (2011).

38. Takahashi, E. J., Hasegawa, H., Nabekawa, Y. & Midorikawa, K. High-throughput, high-damage-threshold broadband beam splitter for high-order harmonics in the extreme-ultraviolet region. *Opt. Lett.* **29,** 507–509 (2004).

39. Kennedy, D. J. & Manson, S. T. Photoionization of the noble gases: cross sections and angular distributions. *Phys. Rev. A* **5,** 227–247 (1972).

40. Lai, C.-J. *et al.* Wavelength scaling of high harmonic generation close to the multiphoton ionization regime. *Phys. Rev. Lett.* **111,** 073901 (2013).

41. Falcão-Filho, E. L., Gkortsas, V. M., Gordon, A. & Kärtner, F. X. Analytic scaling analysis of high harmonic generation conversion efficiency. *Opt. Express* **17,** 11217–11229 (2009).

42. Kazamias, S. *et al.* Pressure-induced phase matching in high-order harmonic generation. *Phys. Rev. A* **83,** 063405 (2011).

43. Yudin, G. & Ivanov, M. Nonadiabatic tunnel ionization: looking inside a laser cycle. *Phys. Rev. A* **64,** 013409 (2001).

44. Gkortsas, V.-M. *et al.* Interplay of multiphoton and tunneling ionization in short-wavelength-driven high-order harmonic generation. *Phys. Rev. A* **84,** 013427 (2011).

45. Graf, J. *et al.* Vacuum space charge effect in laser-based solid-state photoemission spectroscopy. *J. Appl. Phys.* **107,** 014912 (2010).

46. Dakovski, G. L., Li, Y., Durakiewicz, T. & Rodriguez, G. Tunable ultrafast extreme ultraviolet source for time- and angle-resolved photoemission spectroscopy. *Rev. Sci. Instr.* **81,** 073108 (2010).

47. Frassetto, F. *et al.* Single-grating monochromator for extreme-ultraviolet ultrashort pulses. *Opt. Express* **19,** 19169–19181 (2011).

48. Caracciolo, E. *et al.* 28-W, 217 fs solid-state Yb:CAlGdO$_4$ regenerative amplifiers. *Opt. Lett.* **38,** 4131–4133 (2013).

49. Gullikson, E. & Henke, B. CXRO x-ray interactions with matter, http://henke.lbl.gov/optical_constants (2010).

**Acknowledgements.** We thank H. Mashiko and T. Sekikawa for interesting discussions, as well as P. Froemel, S. Stoll, and L. Zeng for help with the experimental setup and amplifier beam characterization. This work was supported by the U.S. Department of Energy, Office of Basic Energy Sciences (DOE BES), Division of Materials Sciences and Engineering under contract DE-AC02-05CH11231, carried out within the Ultrafast Materials Science program at Lawrence Berkeley National Laboratory. S. U. acknowledges a fellowship from the German Academic Exchange Service (DAAD).

**Author Contributions.** H.W. designed and performed the HHG experiments, with assistance from Y.X. and S.U. Moreover, H.W. and R.K. analyzed the results, carried out simulations, and wrote the manuscript. All authors contributed to the discussion of the experimental data and the manuscript.

**Additional Information**

The authors declare no competing financial interests.

**Supplementary Information** is available online at http://dx.doi.org/10.1038/ncomms8459.